%% file: CUOREpos.tex
\documentclass[a4paper,11pt]{article}
\usepackage{pos}

\title{New results from the CUORE experiment}
\ShortTitle{Results from CUORE}

\input{CUOREauthors.tex}

\abstract{The Cryogenic Underground Observatory for Rare Events (CUORE) is the first cryogenic experiment searching for neutrinoless double-beta ($0\nu\beta\beta$) decay that has been able to reach the one-ton scale. The detector, located at the Laboratori Nazionali del Gran Sasso in Italy, consists of an array of 988 TeO$_2$ crystals arranged in a compact cylindrical structure of 19 towers. Following the completion of the detector construction in August 2016, CUORE began its first physics data run in 2017 at a base temperature of about 10\,mK. Following multiple optimization campaigns in 2018, CUORE is currently in stable operating mode. In 2019, CUORE released its 2\textsuperscript{nd} result of the search for $0\nu\beta\beta$ with a TeO$_2$ exposure of 372.5\,kg$\cdot$yr and a median exclusion sensitivity to a $^{130}$Te $0\nu\beta\beta$ decay half-life of $1.7\cdot 10^{25}$\,yr. We find no evidence for $0\nu\beta\beta$ decay and set a 90\% C.I. (credibility interval) Bayesian lower limit of $3.2\cdot 10^{25}$\,yr on the $^{130}$Te $0\nu\beta\beta$ decay half-life. In this work, we present the current status of CUORE's search for $0\nu\beta\beta$, as well as review the detector performance. Finally, we give an update of the CUORE background model and the measurement of the  $^{130}$Te two neutrino double-beta ($2\nu\beta\beta$) decay half-life.}

\FullConference{ICHEP2020\\
                28 July 2020 to 6 August 2020\\
                Prague}

\begin{document}
\maketitle
\section{Introduction}
Experiments on flavor oscillations in solar, atmospheric, accelerator, and reactor neutrinos
have made tremendous progress in pinning down the neutrino mixing angles and oscillation frequencies, providing also the indisputable evidence of the non-vanishing nature of the neutrino mass~\cite{Esteban_2020}. To complete this scenario next-generation experiments are needed to study the mass hierarchy of the neutrino mass eigenstates and the neutrino-antineutrino dichotomy~\cite{Vuilleumier_2018} (\textit{Majorana particle}: $\nu\equiv\overline{\nu}$ or \textit{Dirac particle}: $\nu\not\equiv\overline{\nu}$). One practical way to investigate these open issues is to search for neutrinoless double beta decay ($0\nu\beta\beta$)~\cite{Dolinski_2019}. This is a rare nuclear process not predicted by the Standard Model in which a pair of neutrons inside a nucleus transforms into a pair of protons, with the emission of two electrons: $(A, Z) \rightarrow (A, Z + 2) + 2e^-$. This transition clearly violates the conservation of the number of leptons and its observation would thus demonstrate that the lepton number is not a symmetry of nature. At the same time, ($0\nu\beta\beta$) provides a key tool to study neutrinos by probing whether their nature is that of Majorana particles and providing us with important information on the neutrino absolute mass scale and ordering~\cite{DellOro_2016,DellOro_2019}.
The Cryogenic Underground Observatory for Rare Events~\cite{CUORE_2002,CUORE_2014} (CUORE) is a running experiment built with the primary goal to search for neutrinoless double beta decay in $^{130}$Te by studying the transition: $^{130}\mbox{Te} \rightarrow ^{130}\mbox{Xe} + 2 e^-$. CUORE is located deeply underground ($\sim 3600$\,m.\,w.\,e.) at the Laboratori Nazionali del Gran Sasso, Italy  and is presently taking data. The experiment is expected to collect data for a total of five years of live-time.

\section{CUORE detector}
The CUORE detector consists of a closely-packed array of 988 TeO$_2$ cubic crystals, $5 \times 5 \times 5$\,cm$^3$ in size, arranged in 19 towers, 13 floors each, with 4 crystals per floor supported inside a copper frame. This corresponds to 206\,kg of $^{130}$Te, considering the natural abundance of 34.2\%. Each crystal is equipped with a neutron-transmutation-doped germanium (NTD-Ge) thermistor for the read-out of the temperature signal and a silicon heater used for the thermal gain stabilisation. In order to operate the detector, a large custom cryogenic system~\cite{CUORE_CRYO_2019} was designed and constructed, satisfying very stringent experimental requirements in terms of high cooling power, low noise environment and high radiopurity. This system allows the experiment to maintain the detector array at a stable temperature around 11.8\,mK. After the successful installation of the detector in the summer 2016 the first CUORE cool-down took place between December 2016 and January 2017. After some initial detector optimization campaigns, CUORE has been collecting data since May 2017, and after upgrade efforts in 2018 and 2019 the experiment has been running continuously. The achieved stable conditions allowed continued data taking with minimal onsite activity also during recent lockdowns. CUORE is the culmination of a multi-decade effort to develop low temperature detectors research for neutrino mass studies. Designing, building, and operating CUORE took over a 10-year period. The successful operation of CUORE marked a major milestone in the history of low-temperature detector techniques and opened the way for large calorimetric arrays (tonne-scale) for rare event physics.

\section{Search for neutrinoless double beta decay}
\begin{figure}[!t]
\centering\includegraphics[width=1\textwidth]{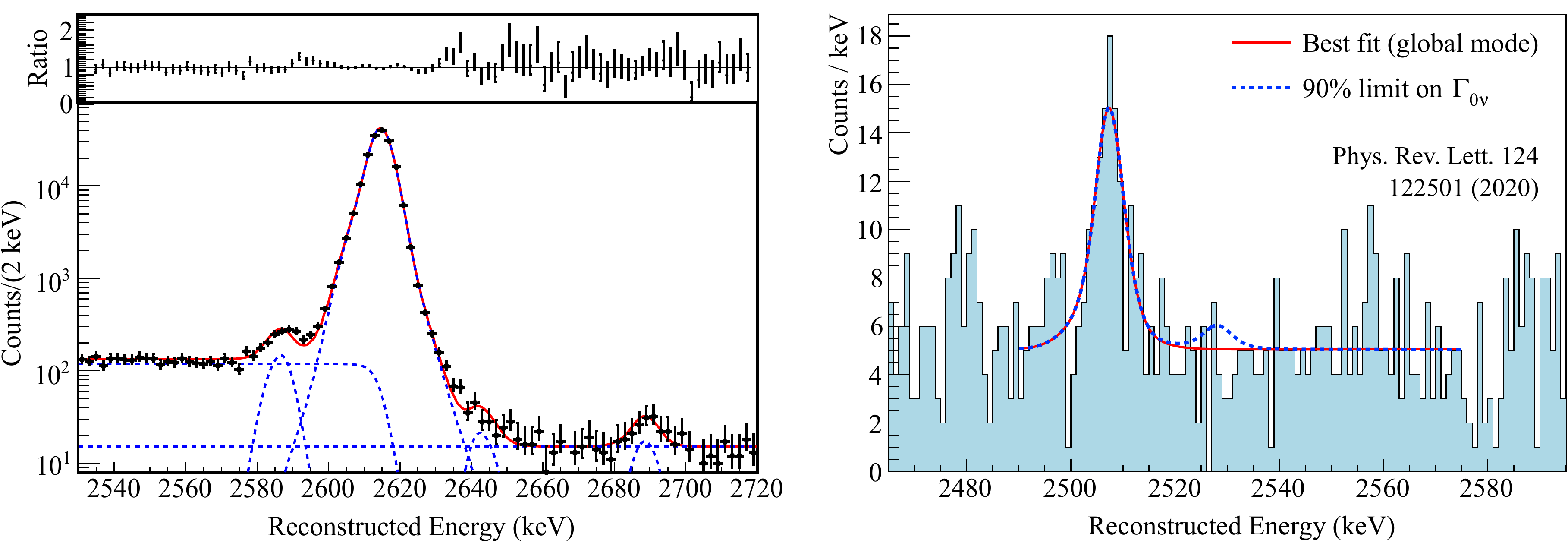}
\caption{(left) Sum of the results of the 19 tower-dependent fits used to evaluate the line shape parameter of each detector in calibration data; (right) Physics spectrum in ROI with the best-fit curve (solid red) and the best fit-curve with the $0\nu\beta\beta$ decay component fixed to the 90\% CI limit (dashed blue)}
\label{fig:spectra}
\end{figure}

CUORE has collected 7 datasets since April 2017. Each dataset includes one-day-long runs for a total of about one month of Physics data, and is started and ended with a calibration measurement. The number of active channels is 984. However, during the analysis, some of these channels were removed for different reasons (e.g. high noise, failure during one or more analysis steps, insufficient statistics collected during calibration, etc). In the end, the analysis was performed with a number of channels that changed dataset by dataset (900 min - 950 max). The average energy resolution at the $\beta\beta$ Q-value, mediated over all the active channels, was $(7.0 \pm 0.4)$\,keV, with an observed improvement during the data collection thanks to the optimization campaign. Since spring 2019, CUORE has been stably collecting data at an average rate of 50\,kg$\cdot$yr/month. During the data-taking campaign between May 2017 and July 2019 CUORE collected a total exposure of 372.5\,kg$\cdot$yr of TeO$_2$ (103.6\,kg$\cdot$yr of $^{130}$Te)~\cite{CUORE_PRL_2020}. 

A blind search for $0\nu\beta\beta$ was performed, and the ROI fit model defined prior to unblinding. The model parameters are the $0\nu\beta\beta$ decay rate, a dataset dependent background index (BI),  the $^{60}$Co sum peak amplitude and its position, which is a free parameter as in the previous analysis~\cite{CUORE_PRL_2017} (figure \ref{fig:spectra}, left). The BIs are dataset dependent, while all other parameters are common to all datasets, including the $^{60}$Co rate, which is scaled by a dataset dependent factor to account for its decay. The line shape parameters for each detector-dataset were estimated with a simultaneous, unbinned extended maximum likelihood (UEML) fit performed on each tower in the energy range (2530 - 2720)\,keV. In particular, all individual detectors were constrained to have the same $0\nu\beta\beta$ decay rate, which we allowed to vary freely in the fit (Figure \ref{fig:spectra}, right). No evidence for $0\nu\beta\beta$ signal was found and the corresponding lower limit on the $0\nu\beta\beta$ half life of $^{130}$Te was set to $T^{0\nu}_{1/2} > 3.2 \cdot 10^{25}$\,yr at 90\% C.L.~\cite{CUORE_PRL_2020}. The average background in the in the $0\nu\beta\beta$ decay region of interest resulted $B = (1.38 \pm 0.07)\cdot 10^{-2}$\,cnts/keV/kg/yr~\cite{CUORE_PRL_2020}, level in line with the expectations. The contribution from $\gamma$ radiation was significantly reduced with respect to CUORE-0, and most of the $\alpha$-induced background was compatible. In the hypothesis that $0\nu\beta\beta$ decay is mediated by light Majorana neutrinos, this results in an upper limit on the effective Majorana mass of 75-350\,meV, depending on the nuclear matrix elements used.

\section{CUORE background}

\begin{figure}[!t]
\centering\includegraphics[width=0.93\textwidth]{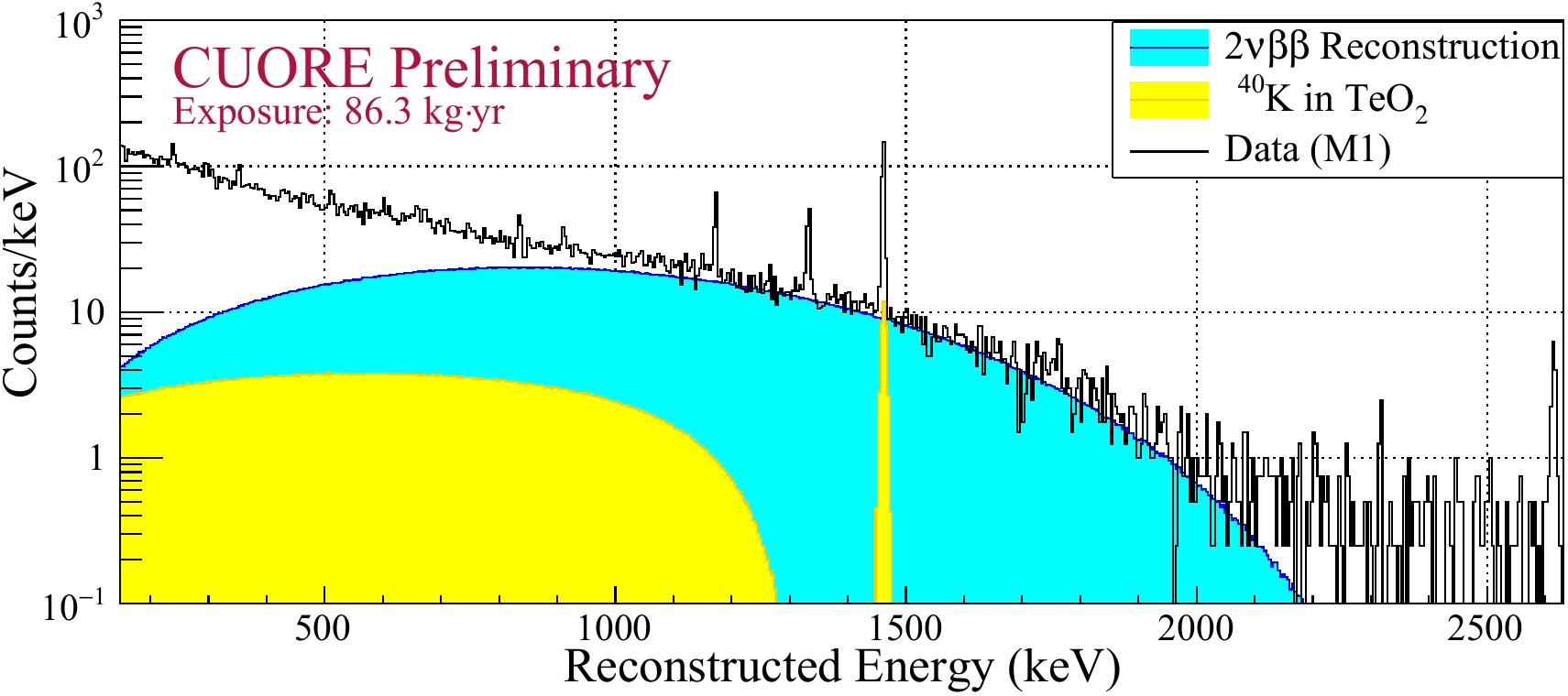}
\caption{CUORE observed spectrum (black) with a reconstruction of the $2\nu\beta\beta$ component (cyan) as well as the $^{40}$K background (yellow). Edited from~\cite{Nutini_2020}.}
\label{fig:bkg}
\end{figure}

In order to systematically study the CUORE radioactive contamination, a background model was developed. This model is able to describe the observed spectrum in terms of contributions from contamination of the materials directly facing the detector, the whole cryogenic setup, and the environmental radioactivity. A set of 62 sources, considering different isotopes and locations, was sufficient to reconstruct the CUORE observed physics data, via a Markov Chain Monte Carlo simultaneous fit across the energy spectra~\cite{Alduino_2017}. Many contaminations were identified and located, and others were still utilized by the fit, but only a limit on their activity was obtained. The model was able to reproduce nearly all the major features of the observed spectra. 

Thanks to developed background model, the background contribution that could be ascribed to the two-neutrino double beta decay $(2\nu\beta\beta)$ of $^{130}$Te was successfully reconstructed (Figure \ref{fig:bkg}). Preliminary results indicated a $(2\nu\beta\beta)$ half-life for $^{130}$Te of $T^{0\nu}_{1/2}=[7.9\pm 0.1 {\small\mbox{(stat)}} \pm 0.2 {\small\mbox{(syst)}}] \cdot 10^{20}$\,year~\cite{Nutini_2020}. This is the world’s most precise measurement for this isotope.

\section{Conclusion}
The CUORE experiment will collect data for a total of five years of live-time. The predicted final sensitivity is $9.0 \cdot 10^{25}$\,yr at 90\% C. L.~\cite{Alduino_2017}. The demonstrated performance and results represent a fundamental step toward the next generation of tonne-scale detector arrays. Starting from the experience, the expertise, and the lessons learned while running CUORE, the CUPID project (CUORE Upgrade with Particle IDentification~\cite{CUPIDInterestGroup_2019}) aims at developing a future large calorimetric $0\nu\beta\beta$ experiment with sensitivity on the half-life of the order of ($10^{27}-10^{28})$\,yr. Thermal detectors are expected to play a central role in the forthcoming future of the search for $0\nu\beta\beta$.

\section*{Acknowledgment}
This work was supported by the Istituto Nazionale di Fisica Nucleare (INFN); the National Science Foundation under Grant Nos. NSF-PHY-0605119, NSF-PHY-0500337, NSF-PHY-0855314, NSF-PHY-0902171, NSF-PHY-0969852, NSF-PHY1307204, and NSF-PHY-1404205; the Alfred P. Sloan Foundation; and Yale University. This material is also based upon work supported by the US Department of Energy (DOE) Office of Science under Contract Nos. DE-AC02-05CH11231 and DE-AC52-07NA27344; and by the DOE Office of Science, Office of Nuclear Physics under Contract Nos. DE-FG02-08ER41551, DE-FG03-00ER41138, DE-SC0011091, and DE-SC0012654. This research used resources of the National Energy Research Scientific Computing Center (NERSC).

\bibliographystyle{JHEP}
\bibliography{CUOREpos}

\end{document}

%% file: CUOREauthors.tex
\author[i,h]{A.~Giachero}
\author[a]{D.~Q.~Adams}
\author[a]{C.~Alduino}
\author[b]{K.~Alfonso}
\author[a]{F.~T.~Avignone~III}
\author[c]{O.~Azzolini}
\author[d]{G.~Bari}
\author[e,f]{F.~Bellini}
\author[g]{G.~Benato}
\author[h]{M.~Biassoni}
% ...

\emailAdd{andrea.giachero@mib.infn.it}

\affiliation[a]{Department of Physics and Astronomy, University of South Carolina, Columbia, SC 29208, USA}
\affiliation[b]{Department of Physics and Astronomy, University of California, Los Angeles, CA 90095, USA}
\affiliation[c]{INFN -- Laboratori Nazionali di Legnaro, Legnaro (Padova) I-35020, Italy}
\affiliation[d]{INFN -- Sezione di Bologna, Bologna I-40127, Italy}
\affiliation[e]{Dipartimento di Fisica, Sapienza Universit\`{a} di Roma, Roma I-00185, Italy}
\affiliation[f]{INFN -- Sezione di Roma, Roma I-00185, Italy}
\affiliation[g]{INFN -- Laboratori Nazionali del Gran Sasso, Assergi (L'Aquila) I-67100, Italy}
\affiliation[h]{INFN -- Sezione di Milano Bicocca, Milano I-20126, Italy}
\affiliation[i]{Dipartimento di Fisica, Universit\`{a} di Milano-Bicocca, Milano I-20126, Italy}
% ...
